\documentclass[10pt]{ismd08}
\usepackage{graphicx}
\usepackage{cite,./mcite}

\begin{document}
\title{Hydrodynamics \& perfect fluids: uniform description of soft observables 
in Au+Au collisions at RHIC \footnote{\,\,\,Supported in part by the Polish Ministry of Science and Higher Education, 
grants N202 153 32/4247 and N202 034 32/0918, and by the U.S. NSF Grant No. PHY-0653432.}}
\author{Wojciech Florkowski$^{1,2}$\protect\footnote{\ \ speaker},
Wojciech Broniowski$^{1,2}$, Mikolaj Chojnacki$^1$, Adam Kisiel$^{3,4}$}
\institute{$^1$The H. Niewodnicza\'nski Institute of Nuclear Physics, \\ Polish Academy of Sciences, PL-31342 Krak\'ow, Poland,\\ 
$^2$Institute of Physics, Jan Kochanowski University, PL-25406~Kielce, Poland,\\ 
$^3$Faculty of Physics, Warsaw University of Technology, PL-00661 Warsaw, Poland, \\
$^4$Department of Physics, Ohio State University, 
1040 Physics Research Building, \\ 191 West Woodruff Ave., 
Columbus, OH 43210, USA }
\maketitle
\begin{abstract}
It is argued that the use of the initial Gaussian energy density profile for hydrodynamics leads to much better uniform description of the RHIC heavy-ion data than the use of the standard initial
condition obtained from the Glauber model. With the modified Gaussian initial conditions we successfully reproduce the $p_T$-spectra, $v_2$, and the pionic HBT radii (including their azimuthal dependence). The emerging consistent picture of hadron production hints that a solution of the long standing RHIC HBT puzzle has been found. 
\end{abstract}

\section{Introduction}
\label{sec:intro}

\begin{figure}[t]
\begin{center}
\includegraphics[angle=0,width=0.45 \textwidth]{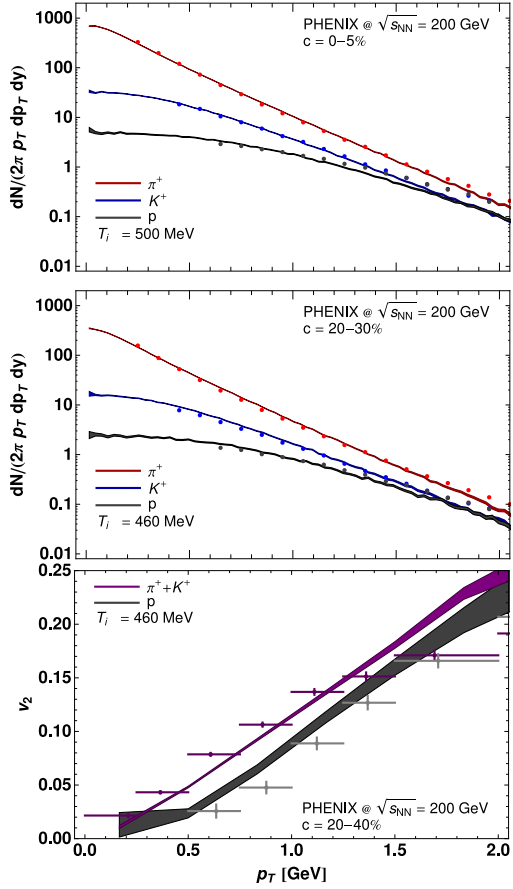}
\end{center}
\caption{The transverse-momentum spectra of pions, kaons and protons for the centrality bin $c$ = 0-5\% (upper panel), $c$ = 20-30\% (middle panel), and the elliptic flow coefficient $v_2$ for  $c$ = 20-40\% (lower panel), plotted as functions of the transverse momentum and compared to the RHIC Au+Au data.}
\label{fig:spv2}
\end{figure}

Relativistic hydrodynamics of the perfect fluid may be considered as the standard framework for the description of the intermediate stages of relativistic heavy-ion collisions \cite{Teaney:2000cw,Hirano:2002ds,Kolb:2003dz,Huovinen:2003fa,Eskola:2005ue,Hama:2005dz,Nonaka:2006yn}. However, despite the clear successes in reproducing the particle transverse-momentum spectra and the elliptic flow coefficient $v_2$, the typical approach based on the relativistic hydrodynamics cannot  reproduce correctly the pion correlation functions. In particular, the ratio of the so called HBT radii $R_{\rm out}$ and $R_{\rm side}$ comes out too large, exceeding the experimentally measured value by about 20-50\%.  Very recently, we have found \cite{Broniowski:2008vp} that the consistent description of the soft hadronic observables may be achieved within the hydrodynamic model if one makes a modification of the initial conditions -- the initial energy profile obtained in most cases from the optical Glauber model should be replaced by the Gaussian profile. 

Our framework consists of the recently developed 2+1 boost-invariant inviscid hydrodynamics  \cite{Chojnacki:2004ec,Chojnacki:2006tv,Chojnacki:2007jc} linked to the statistical-hadronization model THERMINATOR \cite{Kisiel:2005hn}. The initial eccentricity is obtained from the Monte-Carlo Glauber model GLISSANDO \cite{Broniowski:2007nz}. The simulations done with GLISSANDO include the eccentricity fluctuations \cite{Miller:2003kd,Bhalerao:2005mm,Manly:2005zy,Voloshin:2006gz,Andrade:2006yh,Alver:2006wh,Hama:2007dq,Alver:2008zz,Voloshin:2008dg}. In each simulated event the distribution of sources (a mixture of the wounded-nucleon contributions and the binary-collision points) is first rotated to the principal-axes frame and then histogrammed. As a result one obtains the two-dimensional profile that takes into account the fluctuations of the principal axes with respect to the reaction plane. This procedure determines the initial energy distribution in the transverse plane which is parameterized as the two-dimensional Gaussian and used as the initial condition for the hydrodynamics. The main characteristic of the hydrodynamic stage is the use of the realistic equation of state which interpolates between the lattice simulations of full QCD and the hadron-gas model. The final stage of the evolution is described with the help of the Monte-Carlo thermal model THERMINATOR which simulates hadron emission from the freeze-out hypersurface delivered by the hydrodynamic calculation. We assume the single freeze-out scenario \cite{Broniowski:2001we,Broniowski:2002nf} with the universal final temperature $T_f = $145 MeV. Besides the final temperature our model has essentially two additional parameters: the initial temperature $T_i$, fixing the absolute normalization of the spectra, and the initial time for the start of hydrodynamics $\tau_0 =$0.25 fm. Of course, for each centrality class we fix the geometric parameters $a$ and $b$ (i.e., the widths of the initial Gaussian energy distribution) by the GLISSANDO simulations as explained above. 

\begin{figure}[tb]
\begin{center}
\includegraphics[angle=0,width=0.95 \textwidth]{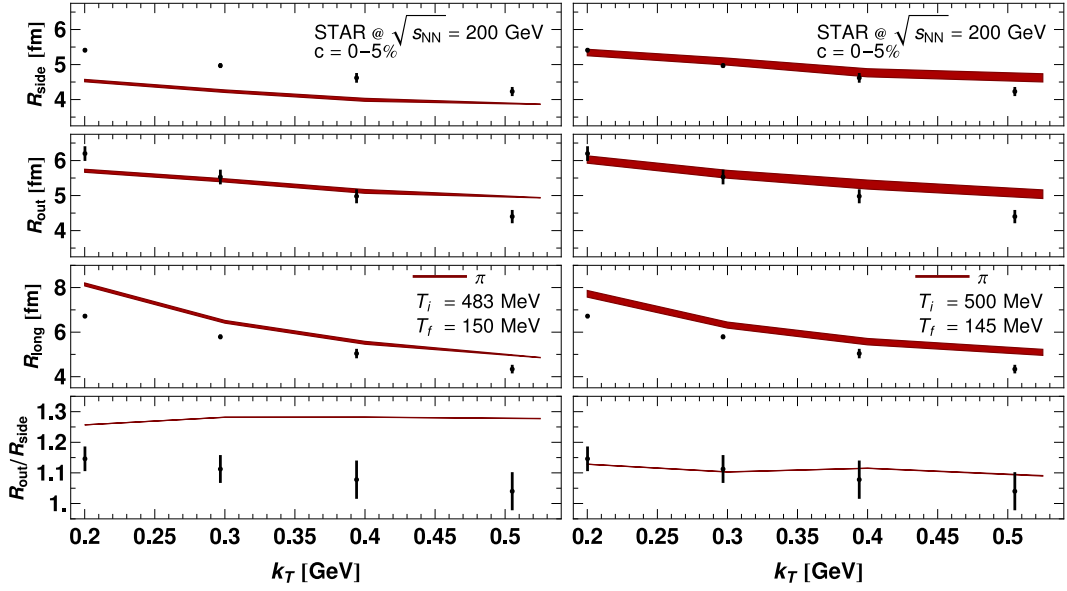}
\end{center}
\caption{ The pion HBT radii $R_{\rm side}$ , $R_{\rm out}$ , $R_{\rm long}$,  and the ratio $R_{\rm out}/R_{\rm side}$ for central collisions, shown as the functions of the average momentum of the pair and compared to the RHIC Au+Au data.
\label{fig:hbt}}
\end{figure}

\section{Results}
\label{sec:results}

In Fig. \ref{fig:spv2} we show our results describing the transverse-momentum spectra of pions, kaons, and protons for the centrality classes $c$ = 0-5\% and $c$ = 20-30\%. Fig. \ref{fig:spv2} presents also our results describing the elliptic flow coefficients $v_2$ for the centrality class  $c$ = 20-40\%, plotted as functions of the transverse momentum. The spectra and the elliptic flow are compared to the RHIC data \cite{Adler:2003cb,Adler:2003kt}. We observe a very good agreement between the model predictions and the data. The small excess of the theoretical proton $v_2$ above the data may be attributed to the lack of rescattering in the final state. 

In Fig. \ref{fig:hbt} we present our results on the pion HBT radii $R_{\rm side}$ , $R_{\rm out}$, $R_{\rm long}$,  and the ratio $R_{\rm out}/R_{\rm side}$ for central collisions, compared to the RHIC data \cite{Adams:2004yc}. The left panel shows our best results obtained with the traditional Glauber initial condition \cite{Chojnacki:2007rq}, while the right panel shows the results obtained with the Gaussian initial condition \cite{Broniowski:2008vp}. One can see that a very good agreement between the data and the theoretical model predictions is achieved in the case where the Gaussian initial condition is used. In particular, the ratio $R_{\rm out}/R_{\rm side}$ is well reproduced. We note that the calculation of the radii does not introduce any extra parameters. All the characteristics of the emitting source were already fixed by the fits to the spectra and $v_2$. 

\begin{figure*}[tb]
\begin{center}
\includegraphics[angle=0,width=0.95 \textwidth]{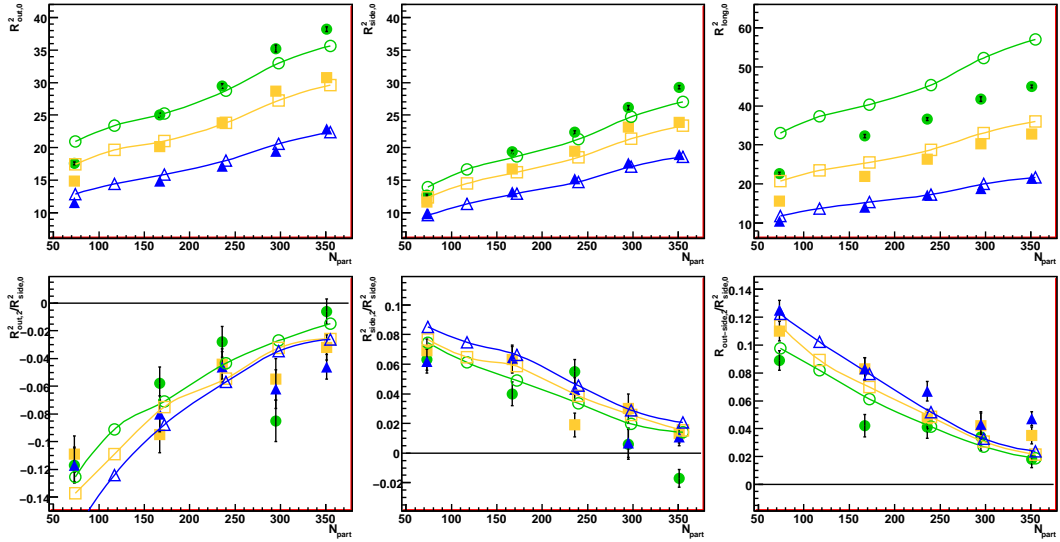}
\end{center}
\vspace{-6.5mm}
\caption{Results for the RHIC HBT radii and their azimuthal oscillations. For
each value of $N_{\rm part}$ on the horizontal axis the
experimental points (filled symbols) and the model results (empty
symbols) are plotted. The points from top to bottom at each plot correspond to
$k_T$ contained in the bins 0.15-0.25~GeV (circles),
0.25-0.35~GeV  (squares), and 0.35-0.6~GeV(triangles). The top panels
show $R^{2}_{\rm out,0}$, $R^{2}_{\rm side,0}$, and $R^{2}_{\rm
long,0}$, the bottom panels show the magnitude of the allowed
oscillations divided conventionally by $R_{\rm side,0}^2$. }
\label{fig:resultsrhic}
\end{figure*}

Finally, in Fig. \ref{fig:resultsrhic} the results describing the azimuthal dependence of the HBT radii are plotted \cite{Kisiel:2008ws}. Here $R^2(\phi) = R_0^2 + 2 R_2^2 \cos(2\phi)$. Again, we observe a very good agreement between the data and our model for different centralities and different average momenta of the pion pairs $k_T$. 
 
\section{Conclusions}
\label{sec:conclusions}

The results presented above indicate that it is possible to achieve a uniform description of the RHIC heavy-ion data collected at the highest beam energies in the soft hadronic sector using the hydrodynamics of perfect fluid with the Gaussian initial condition for the energy density. In particular, it is possible to describe the transverse-momentum spectra and the elliptic flow coefficient $v_2$ simultaneously with two-particle observables such as the HBT correlation radii (the preliminary results show also that the correlations of the non-identical particles are well reproduced in our model). Our finding shows that there exists a solution to the long standing RHIC HBT puzzle understood as the impossibility of the consistent description of the spectra and the HBT radii in a single hydrodynamic approach. 

The use of the Gaussian initial profile leads to a faster development of the initial transverse flow which makes the system evolution shorter. At the same time the transverse size of the system at freeze-out is slightly larger (as compared to the standard Glauber scenario). These two effects put together lead to the desired reduction of the ratio $R_{\rm out}/R_{\rm side}$. Another important effects helping to describe correctly the data consists of the use of the semi-hard equation of state (with no soft point leading to the extended duration of hadronization) and of the adoption of the single freeze-out scenario which also reduces the emission time, hence, decreasing the radius $R_{\rm out}$. 

Of course, the open question remains to find the microscopic mechanism leading to the Gaussian initial conditions. Needless to say, this problem goes far beyond the straightforward application of the hydrodynamics that were discussed here. Alternatively, one may think of other modifications of the initial conditions, such as introducing the initial transverse flow  \cite{Chojnacki:2004ec,Gyulassy:2007zz,Sinyukov:qm08} or separating the system into a thermalized core and an outer mantle/corona consisting of independent $NN$ collisions \cite{Bozek:2005eu,Becattini:2008yn,Bozek:2008zw}. Yet another direction is to study the effects of viscosity \cite{Pratt:2008sz}.

\begin{footnotesize}
\bibliographystyle{ismd08} 
{\raggedright
\bibliography{rrr}
}
\end{footnotesize}
\end{document}